\documentclass[showpacs,twocolumn,nofootinbib]{revtex4}%
\usepackage[colorlinks=true]{hyperref}

\AtBeginDocument{
	\hypersetup{
		urlcolor=cyan,
		citecolor=green,
		linkcolor=red,
		anchorcolor=blue,
	}%
}
\usepackage{amssymb}
\usepackage{amsmath}
\usepackage{amsfonts}
\usepackage{graphicx}
\usepackage{ulem} 
\usepackage{cancel}
\usepackage[usenames]{color}
\usepackage{epstopdf}
\usepackage{extarrows}
\usepackage{fixmath}

\begin{document}

\title[ ]{Reply to ``Comment on `R\'enyi entropy yields artificial biases not in the data and incorrect updating due to the finite-size data'  ''}
\author{$^{1}$Thomas Oikonomou}
\email{thomas.oikonomou@nu.edu.kz}
\author{$^{2}$G. Baris Bagci}
\affiliation{$^{1}$Department of Physics, Nazarbayev University, Nur-Sultan, Kazakhstan}
\affiliation{$^{2}$Department of Physics, Mersin University, Mersin, Turkey}
\keywords{}
\pacs{05.20.-y, 02.50.Tt, 89.70.Cf}

\begin{abstract}
We reply to the Comment by Jizba and Korbel \cite{JKComment} by first pointing out that the Schur-concavity proposed by them falls short of identifying the correct intervals of normalization for the optimum probability distribution even though normalization is a must ingredient in the entropy maximization procedure. Secondly, their treatment of the subset independence axiom requires a modification of the Lagrange multipliers one begins with thereby rendering the optimization less trustworthy. We also explicitly demonstrate that the R\'enyi entropy violates the subset independence axiom and compare it with the Shannon entropy. Thirdly, the new composition rule offered by Jizba and Korbel are shown to yield probability distributions even without a need for the entropy maximization procedure at the expense of creating artificial bias in the data.
\end{abstract}

\eid{ }
\date{\today }
\startpage{1}
\endpage{1}
\maketitle

%

We have recently shown that the R\'enyi entropy violates the Shore-Johnson  (SJ from now on) subset and system independence axioms  \cite{OikGBB1,SJ,note}. Jizba and Korbel (JK hereafter) hold different views in this regard  \cite{JK}. Therefore, we consider each SJ axiom below in items apart from the second one i.e. the invariance axiom on which we have a consensus with JK.

\textit{1. Uniqueness axiom}: In our assessment of this axiom, we have used the concavity as a criterion which led us to the $q$ interval being $(0,1)$. Inspecting the R\'enyi maximum distribution one can see that for a variable interval $x\in(x_{\min},x_{\max})$ with a proper $x_{\min}$ the distribution is normalizable for $0<q<1+(x_{\max})^{-1}$. Apparently, in this inequality we identify also $q$-values greater than the unity. The Schur-concavity that JK used as uniqueness criterion indeed allows these values. However, there is a major drawback in this consideration, namely JK in their entire derivation assumed that $q$ and $x$ are mathematically independent quantities. As can be seen from the inequality above, this only holds when $q\in(0,1)$, since for $q>1$ the deformation parameter carries information about $x$. Thus, the $q>1$ values are irrelevant and must be discarded. Accordingly, the Schur-concavity criterion, either at the pre-maximization or at the post-maximization state, will be reduced to the ordinary concavity criterion. From a different point of view, the inequality above unveils that when $q>1$, the $q$ value is dictated by the finite size of the data, which agrees with the title of our work. Since this happens anyway, the interval of interest to be explored is for $q\in(0,1)$.

\textit{3. Subset independence axiom}:  JK, contrary to our opinion, argue that the R\'enyi entropy satisfies this axiom. The crux of their argument can be traced back to the idea that the maximization of $f(\sum_i g(p_i))$ should yield the same result as $\sum_i g(p_i)$. This certainly looks correct \textit{prima facie} when one considers the entropy maximization procedure as \textbf{already} carried out i.e. at a post-maximization stage. However, as explicitly pointed out in Ref. \cite{Presse1}, the SJ axioms are concerned with the pre-maximization stage and they are about choosing the functional which is only later to be maximized to draw consistent inferences from the data. Since SJ axioms are directly related to the pre-maximization in order to close the door to a possible erroneous inference procedure, these axioms yield different judgements facing the entropies that are closely related to one another. A case in point is to consider the Tsallis and R\'enyi entropies. Although they are monotonically related to one another, the use of SJ axioms shows that the Tsallis entropy only violates the system independence axiom whereas the R\'enyi entropy violates both the subset independence and system independence axioms with ordinary linear averaged constraints \cite{OikGBB1,Presse2}. This difference between the two occurs not because of additivity versus non-additivity, since SJ axioms are general enough. The reason is that SJ axioms consider the pre-maximization stage and check what can be consistently maximized or not (and in which interval of admissible parameters) beforehand. Otherwise, one can surely maximize the R\'enyi entropy and obtain the concomitant probability distribution. SJ axioms just warn us that we should not trust this distribution for consistent inferences.

Second related issue can be explicitly observed from the treatment of the subset independence by JK. Note that for their discussion in Eqs. (7)-(14), JK are forced to change the Lagrange multipliers in order to achieve their statement about the equivalence $f(\sum_i g(p_i)) \sim \sum_i g(p_i)$ to confirm the subset independence for the R\'enyi entropy.  This is erroneous, since one can easily verify that the maximization of $f(\sum_i g(p_i))$ characterizes a different maximum state rather than the maximization of $\sum_i g(p_i)$, i.e., $\beta = \frac{\partial \sum_i g(p_i)}{\partial U}$ while $\beta\neq \frac{\partial f(\sum_i g(p_i))}{\partial U}$ in their notation \cite{Karabulut,OikGBB2}. It is also worth emphasis that this move already shows that there is something wrong with the maximization procedure carried out in this manner, since it yields the infamous physical temperature problem plaguing the field. In other words, due to this manipulation by JK, the Lagrange multiplier $\beta$ is not inverse temperature and even worse does not yield the same value as the one obtained from the Clausius relation \cite{Abe}. Note that one also looses the connection with thermodynamics as a result of such an unjustified entropy maximization procedure (see Ref. \cite{Plastino} for a nice exposition of such a case).

\begin{figure}[!h]
	\centering
	\includegraphics[keepaspectratio,width=7cm]{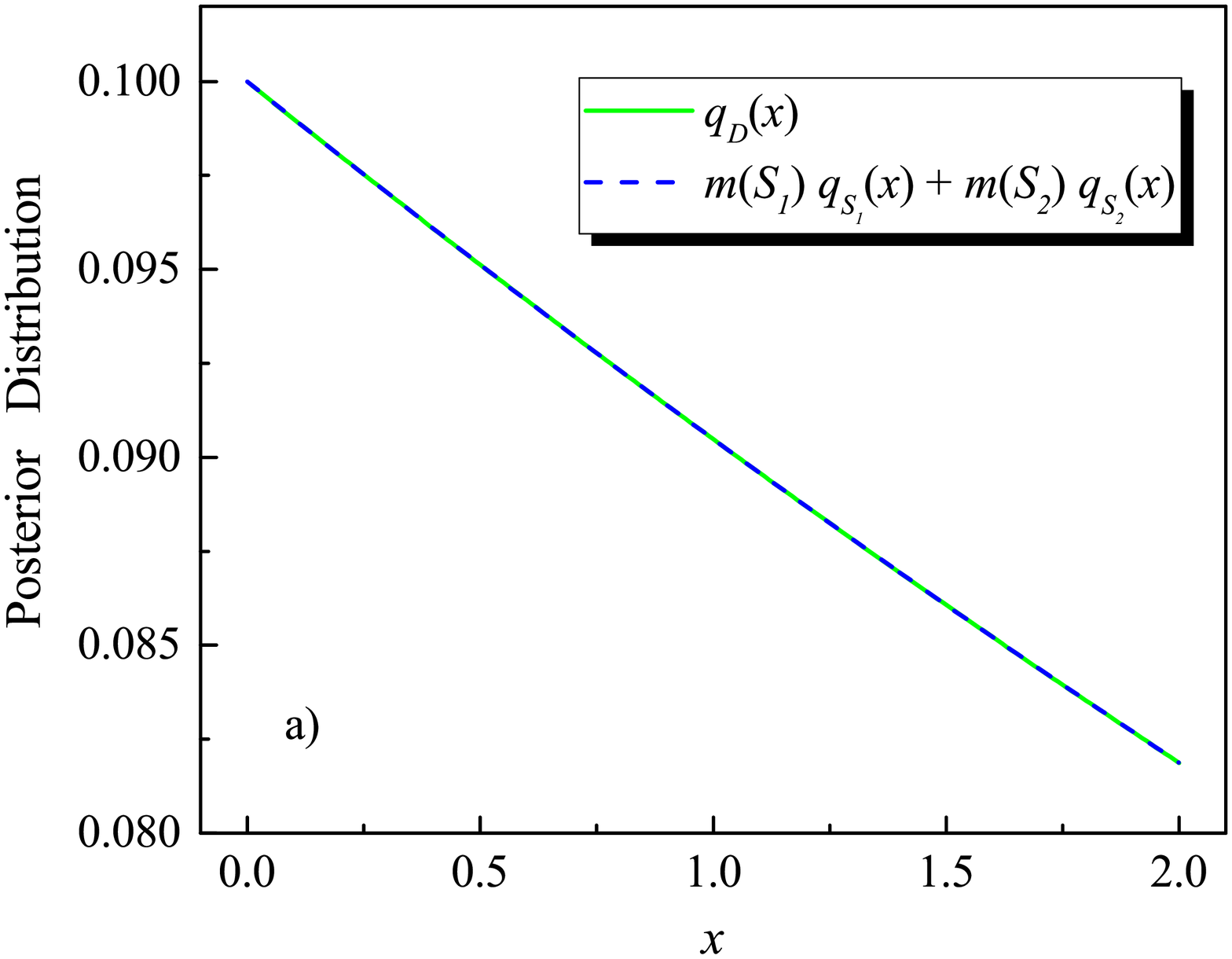}
	\quad
	\includegraphics[keepaspectratio,width=6.7cm]{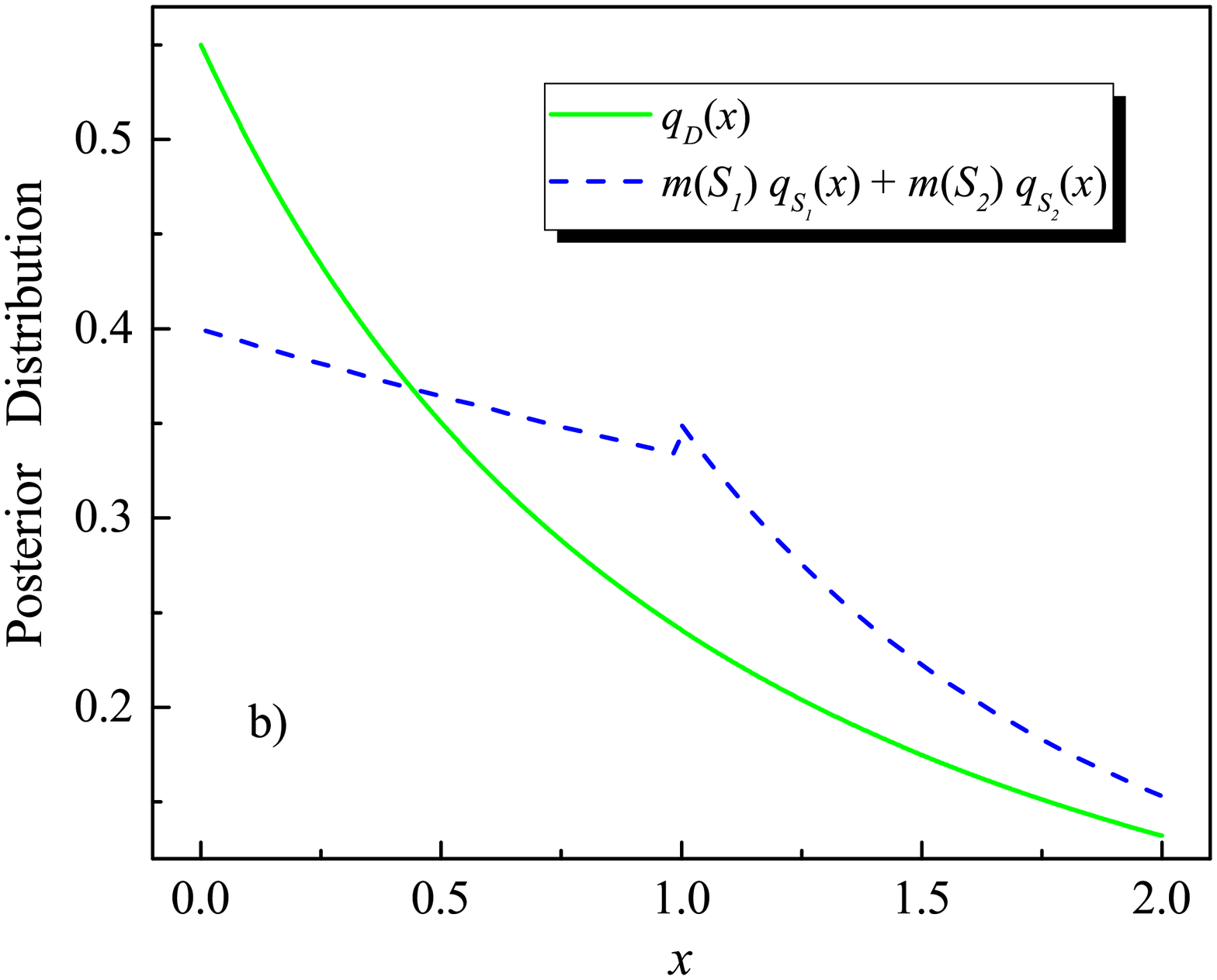}
	\caption{Posterior (maximum) probability distributions for the a) Shannon and b) R\'enyi entropies. In both graphs the green solid and the blue dashed lines are the posterior distributions in the l.h.s. and the r.h.s. of Eq. (\ref{SubSetInd}), respectively. Fig. a) shows the fulfillment of the subset independence for the Shannon entropy, while Fig. b) shows the violation of the former axiom for the R\'enyi entropy.}
	\label{fig1}
\end{figure}

Thirdly, one can easily show from their very Eq. (14), by eliminating the Lagrange multiplier $\alpha$ invoking the maximization conditions, that the maximum distribution contains a cross-term, i.e.  $U$, whose values depend on the particular subset under scrutiny therefore violating the subset independence. We also provide  a simple example, based on the original SJ criterion for continuous variables describing  the subset independence axiom, to numerically demonstrate this violation in the R\'enyi case and compare it with the Shannon case.
According to SJ, for the total set $D=S_1\cup S_2$ with $S_1\cap S_2= \emptyset$, the subset independence axiom is satisfied when (see Eqs. (13)-(14)  in Ref. \cite{SJ})
\begin{eqnarray}\label{SubSetInd}
q_D(x)=m(S_1) q_{S_1}(x)+m(S_2) q_{S_2}(x)\,,
\end{eqnarray}
where $q_D(x), q_{S_1}(x)$ and $q_{S_2}(x)$ are the optimized distributions in the respective set, and
\begin{eqnarray}\label{Coeff}
m(S_i)=\int_{x\in S_i} q_D(x)\mathrm{d}x\quad\Rightarrow\quad  m(S_1)+m(S_2)=1\,.
\end{eqnarray}
Consider then $D=[0,\infty)=S_1\cup S_2=[0,1)\cup[1,\infty)$. 
The Shannon posterior distributions are calculated from the MaxEnt procedure to be
\begin{eqnarray}\label{S_Post}
\nonumber
&& q_D(x)=\beta e^{-\beta x}\,, \qquad  q_{S_1}(x)=\frac{\beta}{1-e^{-\beta}}\,e^{-\beta x}\,,\\
&& q_{S_2}(x)=\beta\,e^{\beta(1-x)}
\end{eqnarray}
with the coefficients determined from Eq. (\ref{Coeff}) above as
\begin{eqnarray}\label{S_Coeff}
m(S_1)=1-e^{-\beta}\,,\qquad m(S_2)=e^{-\beta}\,.
\end{eqnarray}
Substituting then Eqs. (\ref{S_Post}) and (\ref{S_Coeff}) into Eq. (\ref{SubSetInd}) we verify the fulfillment of the subset independence axiom  for the Shannon case.
The R\'enyi posterior distributions, on the other hand, are given by
\begin{eqnarray}\label{R_Post}
\nonumber
&&q_D(x)=\frac{\beta e_q^{-\beta(x-U)}}{\left[e_q^{\beta U}\right]^q}\,,\;
q_{S_1}(x)=\frac{\beta e_q^{-\beta(x-U_1)}}{\left[e_q^{\beta U_1}\right]^q-\left[e_q^{\beta(U_1-1)}\right]^q}\,, \\
&&q_{S_2}(x)=\frac{\beta e_q^{-\beta(x-U_2)}}{\left[e_q^{\beta(U_2-1)}\right]^q}
\end{eqnarray}
where we introduced the notation $e_q^x:=[1+(q-1)q^{-1}\,x]^{1/(q-1)}$ and $\{U,U_1,U_2\}$ are the mean values in the respective set $\{D,S_1,S_2\}$. Also the probability normalization in each set requires $q<1$ with $q>\beta U(1+\beta U)^{-1}$, $q>\beta U_1(1+\beta U_1)^{-1}$ and $q>\beta U_2(1+\beta U_2)^{-1}$. The coefficients $m(S_i)$ are calculated again by means of Eq. (\ref{Coeff}) as
\begin{eqnarray}\label{R_Coeff}
m(S_1)=1-\left[ \frac{e_q^{\beta(U-1)}}{e_q^{\beta U}}\right]^{q}\,,\quad 
m(S_2)=\left[ \frac{e_q^{\beta(U-1)}}{e_q^{\beta U}}\right]^{q}\,.
\end{eqnarray}
As can be immediately seen here, the coefficients depend on the mean value of the entire set $D$ instead of $S_i$ which causes the violation of the subset independence. Indeed, substituting Eqs. (\ref{R_Post}) and (\ref{R_Coeff}) into Eq. (\ref{SubSetInd}) we verify that the latter is not satisfied. For the reader's convenience we demonstrate graphically our results. In Figs. \ref{fig1}a) and \ref{fig1}b) we plot the l.h.s and r.h.s. of Eq. (\ref{SubSetInd}) for the Shannon and R\'enyi entropies, respectively, for the randomly chosen values $\beta=0.1$ and $q=0.55$ (normalization conditions taken into account).

Correctly then, the Livesey-Skilling criterion unveils that only trace-form entropies satisfy the subset independence axiom. We note here that the  Livesey-Skilling criterion for capturing the subset independence is more accurate than the one originally presented by Shore and Johnson. To show this, however, lies outside the scope of this short reply and will be demonstrated by the current authors separately.\\

\textit{4. System independence axiom}: First of all, JK in this part of their comment compare our discussion  with their modified SJ axiom in Ref. \cite{JK} (``... \textit{strong system independence} is added to the SJ desiderata'') and not with the original SJ axiom in \cite{SJ}. In this sense their comment is irrelevant for our discussion in \cite{OikGBB1}. 

Albeit, let us closely inspect the modified axiom introduced by JK. The main move of JK is to introduce a probability composition rule different from the one adopted by SJ. As a result, instead of the original composition rule $p_{ij}=u_iv_j$, one now has $g(p_{ij})=g(u_iv_j)=g(u_i)g(v_j)$ (supplemental material of \cite{JK}, p.3). However, once one introduces this novel composition rule, then the solution is given in advance and independent from the maximization procedure, namely $g(x)\sim x^{q}$ \cite{Hanel}. In other words, through the modification proposed by JK, the entropy maximization procedure becomes redundant. Then, one naturally asks what the use of the SJ axioms could be in the first place if they would be deemed redundant anyway by modifying only one of the axioms. Note also that one can obtain any entropic structure one wishes by such modifications of this axiom completely bypassing the entropy maximization procedure. 

Lastly, as already noted in Ref. \cite{Presse2}, different composition rules, e.g. generalized $q$-products, introduce biases not warranted in the data at all. 
We can understand this in the following simple way. At the maximum state we must have $\sum_{i,j}p_{ij}a_i=\overline{a}=\sum_{i}u_i a_i$, or equivalently $\sum_{i,j}a_i(p_{ij}-u_iv_j)=0$ (the same of course holds for the second system too). Apparently, the fulfillment of this equation is always met only when $p_{ij}=u_i v_j$. Therefore, the use of the multiplicative joint probability composition rule at the pre-maximization state ensures the accuracy of the SJ inference procedure.

%
%
%

Before concluding, JK correctly emphasize that the scope of the original SJ axioms is linear constraints and the use of the escort averages (or any other averaging procedure) remains an open problem. Since JK did not present any explicit inconsistency in our treatment, we provide our view on this issue here. SJ axioms are conceptually general enough in the sense that the violation of the second axiom for example dooms an entropy measure to the discrete uses only independent of the averaging scheme. In fact, it is important to remember that the entropy measures themselves are obtained from specific averaging procedures. In this regard, note that the R\'enyi entropy itself is obtained from the exponential averaging of the information gain whereas the Shannon entropy is formed by the linear average of the very same information gain. Therefore, the main issue is not the averaging procedure inherent in the entropy expressions but the constraints as duly noted by JK. However, constraints only enter into the SJ axioms when the functional is under scrutiny. For example, when one checks the second axiom above, we do not look for any averaging procedure in general, but try to find a consistent continuous generalization of the discrete entropy expression \cite{OikGBB3}. In other words, the conceptual message of the  SJ axioms allows us to choose the appropriate averaging procedure for a new entropy measure so that the entropy measure can be used to draw consistent inferences from the data. For linearly averaged constraints, however, we uniquely have the Shannon entropy.

To sum up, contrary to the criticisms of JK, we showed that i) invoking Schur-concavity as a criterion for the uniqueness of the maximization solution, in contrast to the ordinary entropy concavity, may violate the probability normalization condition, ii) the Livesey-Skilling criterion correctly describes the subset independence axiom, allowing only trace form entropies for a self-consistent inference procedure, and iii) the modified JK subsystem independence axiom renders the maximization procedure redundant.

\begin{acknowledgments}
T.O. acknowledges support by the state-targeted program ``Center of Excellence for Fundamental and Applied Physics" (BR05236454) by the Ministry of Education and Science of the Republic of Kazakhstan, and by the ORAU grant entitled ``Casimir light as a probe of vacuum fluctuation simplification" with PN 17098. G.B.B. acknowledges support from Mersin University under the project number 2018-3-AP5-3093.
\end{acknowledgments}



\begin{thebibliography}{000}
\bibitem{note} Our result is valid whenever one optimizes the R\'enyi entropy. Hence, it does not impose any restrictions on the quantum information theoretical applications of this entropy as long as entropy optimization is not included.
\end{thebibliography}
\end{document}